\documentclass[10pt,aps,longbibliography,prd,superscriptaddress,twocolumn,amsmath,amssymb,floatfix,
nofootinbib 
]{revtex4-1}   
           
\usepackage[dvips]{graphics}
\usepackage{bm}
\usepackage{epsfig}
\usepackage{enumerate}
\usepackage{subfigure}
\usepackage{color}
\usepackage{graphicx} 
\usepackage[dvipsnames]{xcolor} 
\usepackage{graphicx}
\usepackage[colorlinks,citecolor=blue,linkcolor=blue,urlcolor=blue,plainpages=false,pdfpagelabels]{hyperref}

\newcommand{\la}{{\langle}}
\newcommand{\ra}{{\rangle}}
\usepackage{tcolorbox}
\usepackage{bbold}

\newcommand{\beq}{\begin{equation}}
\newcommand{\eeq}{\end{equation}}
\newcommand{\bea}{\begin{eqnarray}}
\newcommand{\eea}{\end{eqnarray}}

\newcommand{\g}{g}

\usepackage{multibib}
\newcites{supp}{Supplementary References}

\begin{document} 
\title{Unconventional superconductivity with preformed pairs in twisted bilayer graphene} 

\author{SK Firoz Islam}
\affiliation{Department of Applied Physics, Aalto University, P.~O.~Box 15100, FI-00076 AALTO, Finland}

\author{A. Yu. Zyuzin}
\affiliation{Ioffe Physical--Technical Institute,~194021 St.~Petersburg, Russia}

\author{Alexander A. Zyuzin}
\affiliation{Department of Applied Physics, Aalto University, P.~O.~Box 15100, FI-00076 AALTO, Finland}

\begin{abstract}
We present a theory of superconductivity in magic-angle twisted bilayer graphene and analyze the superconducting phase diagram in presence of the magnetic field. 
Namely, we consider a model of a granular array hosting localized states, which are hybridized via the delocalized fermions in the inter-grain regions. We study a strong coupling situation when the interactions lead to an incoherent state with preformed Cooper pairs inside the grains. 
The Andreev scattering among different grains manifests itself through the global phase-coherent superconducting state at lower temperatures. 
We demonstrate that a new phase transition between the preformed Cooper pairing state and the Larkin-Ovchinnikov-Fulde-Ferrell state might be induced by the spin pair-breaking effect of in-plane magnetic field. 
The upper critical magnetic field is shown to be enhanced in the strong coupling case. 
\end{abstract}  
\maketitle

Bilayer graphene with a small twist angle between the layers, twisted bilayer graphene (TBG), forms a triangular moire superlattice   \cite{PhysRevLett.99.256802}. It might be seen as an array of domains (grains) that can be approximated by the so-called AA-stacking configuration of the two graphene layers. The domains are spatially separated by the regions which can be described by the Bernal stacking configuration.   
Several scanning tunnelling microscope experiments have confirmed the moire pattern in TBG \cite{PhysRevB.92.155409,PhysRevLett.109.196802,Nature_109.2009}. 
  
Early theoretical studies revealed that low-energy electronic states of TBG obey the Dirac equation with the renormalized Fermi velocity \cite{Nano_Lett,PhysRevB.86.125413,PhysRevB.82.121407,Bistritzer12233}. The electronic properties of TBG are extremely  sensitive to the small twist angle. Most importantly, the Fermi velocity can be strongly suppressed at several twist angles (known as magic angles), leading to almost dispersionless flat bands \cite{Nano_Lett,PhysRevB.86.125413,PhysRevB.82.121407,Bistritzer12233}. The existence of the quasi-flat band has been confirmed in experiments \cite{Nature_109.2009} by the observation of the local density of states peak inside the {\rm AA}-domains.
Moreover, a direct imaging technique has also recently unveiled the flat bands in TBG \cite{Lisi_2020}. 
                                    
Recently, the experimental observations of the two-dimensional (2D) superconductivity at a magic angle TBG have resurgent one of the most debated topics of superconductivity in flat-band materials \cite{Cao2018}. Prior to these experiments, the possibility of flat-band stimulated Cooper pairing with transition temperature linearly dependent on the interaction strength was considered, see for a review Ref.~[\onlinecite{heikkila2016}]. Moreover, an interesting aspect of $2$D superconductor is that the impacts of an in-plane magnetic field can be exclusively considered through the Zeeman pair-breaking effect with strongly suppressed interference from the orbital contributions. It leads to some interesting phenomena. For example, the in-plane magnetic field can enhance the transition temperature in some $2$D superconductors \cite{critical_field,review_nature}. 
It is also known to cause the formation of inhomogeneous Larkin-Ovchinnikov-Fulde-Ferrell (LOFF) type of pairing, for a review  \cite{Hiroshi_review, RevModPhys.76.263}.   

The very recent experiment observed that with the decrease of temperature a pseudogap emerges prior to the transition to the superconductivity in TBG \cite{Oh_2021}. Such pseudogap state could be the result of the formation of incoherent short-ranged Cooper pairs which are formed out of the localized electrons inside the ${\rm AA}$ domains and does not exhibit phase-coherent superconductivity. The model of preformed Cooper pairs was previously discussed in the context of high-temperature \cite{JLTP_Schmitt, Yanase, Geshkenbein}, disordered amorphous superconductors \cite{Nature_Sacepe_first, Nature_Sacepe_second}, and recently in the context of semiconductors and flat-band semimetals \cite{Ayu_zyuzin, Zyuzin_Zyuzin}. For details see the review in Ref.~\cite{Hashimoto_review, Nature_Sacepe_Feigelman}.   
  
In this paper, we propose the model of superconductivity in TBG. First, we analyze the preformed Cooper pairing of localized fermions inside the {\rm AA}-domains in the strong coupling limit. Secondly, we obtain the transition temperature to the phase-coherent state, driven by the inter-domain Andreev coupling. We then study the Zeeman spin splitting effect on the phase diagram focusing on the transition between preformed Cooper pair and LOFF states. 
Finally, we comment that the critical value of the out-of-plane magnetic field which destroys the phase-coherent superconductivity can be increased in the strong coupling case.

\begin{figure}[t!]
\centering
\begin{minipage}[t!]{.5\textwidth}
 \hspace{-0.14cm}{ \includegraphics[width=.5\textwidth, height=4cm]{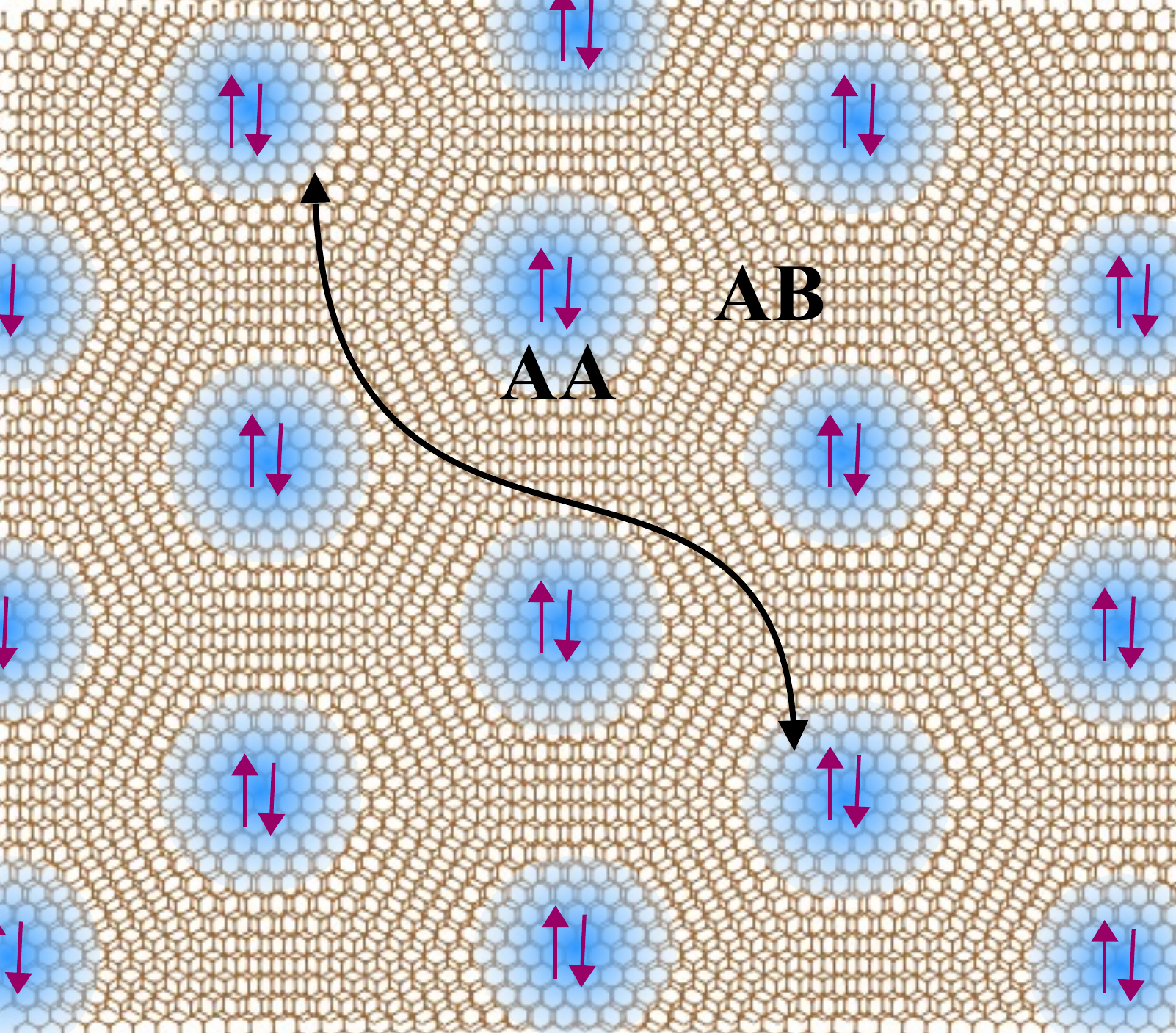}}
  \hspace{-0.07cm}{ \includegraphics[width=.41\textwidth, height=2.3cm]{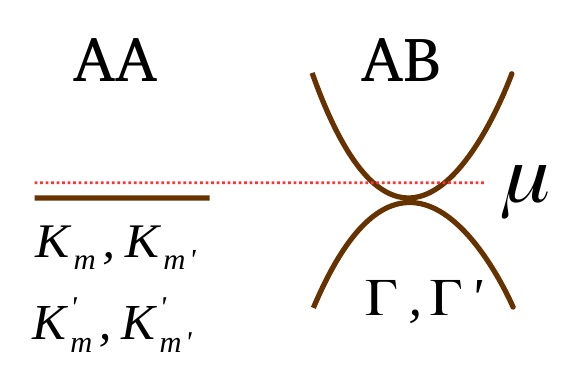}}
\end{minipage}  
\caption{(left) Moire superlattice in twisted bilayer graphene is depicted, where blue spots define {\rm AA}-stacked domains and the rest of bilayer is in {\rm AB}-stacked configuration. The curly arrow denotes long-range Andreev coupling between two domains. (right) The energy spectra in these two regions are schematically shown. The flat-band states are restricted inside the {\rm AA}-domain. The parabolic spectrum describes delocalized states inside {\rm AB}-region. The dashed red line denotes the chemical potential $\mu$.
}
\label{AA_BA}
\end{figure}

\emph{Model of TBG}.
The moire pattern, which is obtained by superposing two graphene layers, consists of two distinct atomic stacking regions, see Fig. (\ref{AA_BA}).   
In the AA-stacked domain every atom in the top layer is positioned approximately above the atom in the bottom layer. The surrounding inter-domain region can be approximated by the AB(BA)-stacking configuration,  
in which each A(B) atom in the top layer is placed above the B(A) atom in the bottom layer. The size of the AA-domains and the distance between them are determined by the twist angle \cite{Nano_Lett, PhysRevB.86.125413}. It was also pointed out that localization inside the ${\rm AA}$-domains can be thought of as a confinement effect due to a hard-wall potential, leading to a quantization condition for momentum \cite{Nano_Lett,PhysRevB.86.125413}.

As for momentum space, the band structure of each isolated layer hosts two non-equivalent valleys at low energy (and at momenta denoted as ${\rm K}$ and ${\rm K'}$). Turning on the inter-layer tunneling in AA-stacked bilayer, the small twist between layers leads to the formation of moire-superlattice of shrunk {\rm AA} domains which are separated by the newly emerged {\rm AB/BA} region. Such twist splits up each valley into two moire valleys i.e., ${\rm K}\rightarrow ({\rm K}_m,{\rm K}_{m'})$ and ${\rm K'}\rightarrow ({\rm K'}_m,{\rm K'}_{m'})$ in the moire-Brillouin zone (mBZ) \cite{Bistritzer12233}. At magic angles, the energy spectrum near these points becomes flat, which promotes localization of electrons inside the {\rm AA}-regions due to nearly zero group velocity.

To proceed, we adopt a model of coexistence of localized and delocalized charge carriers in magic angle TBG \cite{song2021matbg}.
The flat-band reveals itself via the fermions, which are localized inside the {\rm AA}-domains. The delocalized fermions in the ${\rm AB}/{\rm BA}$-stacked regions obey the parabolic dispersion typical for the Bernal stacked bilayer graphene \cite{PhysRevLett.119.107201}. 
Here the effective mass inside the {\rm AB}-region is less sensitive to the twist-angle \cite{Nature_109.2009}. In our model, the coupling between the ${\rm AA}$-domains is provided by delocalized fermions in the ${\rm AB}/{\rm BA}$-regions \cite{song2021matbg}. 

We model magic-angle TBG subject to the magnetic field by the Hamiltonian $\mathcal{H}=\mathcal{H}_{\rm AA}+\mathcal{H}_{\rm AB}+\mathcal{H}_{\rm{T}}+\mathcal{H}_{\rm{int}}$, where each term will be described in what follows. The Hamiltonian describing the {\rm AA}-region is given by
\begin{equation} 
 \mathcal{H}_{\rm AA}=\int _{\bf r}\sum_{i;s,\xi} c_{i;s,\xi} ^{\dagger}({\bf r}) (s\omega_z - \mu) c_{i;s,\xi}({\bf r}),
\end{equation}
where $s=\pm$ denotes two spin-projections and two moire-valleys ${\rm K}_m, {\rm K'}_{m'}$ are denoted by $\xi=\pm$. The fermion creation operator in sublattice space is written as
$c_{i;s,\xi} = [c_{i; s,\xi}^A~c_{i;s,\xi}^B{~}]^{\mathrm{T}}$.
Note that there is also another pair of moire-valleys (${\rm K}_{m'}, {\rm K'}_{m}$) that only increases the size of the vector without any significant impact on spectrum and the further discussion, hence we suppress it.
The chemical potential $\mu$ and the Zeeman energy $\omega_z$ don't depend on domain index $i$. The contribution of the orbital effect of the magnetic field is strongly suppressed at the magic angle. We use short hand notation for the integral $\int {d\bf r}(..) \equiv \int_{\bf r}(..)$, which runs over the area $S_{\mathrm{G}}$ of the ${\rm AA}$-domain. We will be using $\hbar=k_{\mathrm{B}}=1$ units throughout the paper.

We attribute delocalized electrons in ${\rm AB}$-stacked region to the states near the $(\Gamma,\Gamma')$ points of two mBZ denoted by same index $\xi =\pm$ as above \cite{song2021matbg}. These $(\Gamma,\Gamma')$ electrons can be modelled by the Hamiltonian of Bernal-stacked bilayer graphene as
\begin{eqnarray}\nonumber
 \mathcal{H}_{\rm AB} &=&\int _{\mathbf{R}} \sum_{s,\xi}d^{\dagger}_{s,\xi}(\mathbf{R}) \bigg\{s\omega_z -\mu +\frac{1}{2m}\big[
(p_x^2-p_y^2)\sigma_x \\
 &+& \xi(p_xp_y+p_yp_x)\sigma_y
\big]\bigg\}d_{s,\xi}(\mathbf{R}).
\end{eqnarray}
The fermion operator in this region is written as $d_{s,\xi} = [d_{s,\xi}^A~d_{s,\xi}^B{~}]^{\mathrm{T}}$. Here the Pauli matrices $\sigma_{x,y}$ act in sublattice space, $m$ is the effective mass, 
operator ${\bf p} = -i\boldsymbol{\nabla}_{R} -(e/c){\bf A}(\mathbf{R})$ takes into account the orbital effect of magnetic field, where ${\bf A}(\mathbf{R})$ is the magnetic vector potential and $e<0$ is the electronic charge.
 The Zeeman energy is taken to be the same as in ${\rm AA}$-domains. 

The hybridization between ${\rm AA}$-domain with it's surrounding ${\rm AB}$-region is treated within the tunnel model approach. The fermionic states from the moire-valley ${\rm K}_{m}({\rm K'}_{m'})$ of {\rm AA}-region hybridize with the states of $\Gamma(\Gamma')$ point of the {\rm AB}-region. We neglect hybridization between ${\rm K}_{m}({\rm K'}_{m'})$ and $\Gamma'(\Gamma)$ which require large momentum shift. The Hamiltonian describing spin, valley, and sublattice conserving hybridization processes is given by 
\begin{equation}\label{Hybrid}
 \mathcal{H}_{\mathrm{T}}=t\int_{{\bf r},\mathbf{R}}\sum_{i;s,\xi}\left[d^{\dagger}_{s,\xi}( \mathbf{R}) c_{i;s,\xi}({\bf r})+{\rm h.c} \right].
\end{equation}
For the estimation of energy parameter $t$ see Supplemental Materials (SM) Ref.~\cite{supple}.

The superconductivity in TBG was reported at the vicinity of a magic angle suggesting that the flat-band localized electrons might play a significant role in the Cooper pair preformation \cite{Cao2018}. Motivated by the experiments, we consider attractive electron-electron pairing interaction inside the ${\rm AA}$-domains and neglect it in the {\rm AB}-region. We consider the simplest phonon-mediated $s$-wave intra-sublattice pairing channel, in which a Cooper pair is composed of electrons from two moire-valleys (${\rm K}_m, {\rm K'}_{m'}$) symmetric under the time reversal \cite{Bernevig2019, Heikkila_meanfield}. Our discussion may be further extended to the case of d-wave pairing channel, which becomes favourable in the case of strong on-site Coulomb repulsion \cite{PhysRevLett.121.257001}.
The pairing interaction in the {\rm AA}-domains is described by
\begin{eqnarray}
 \mathcal{H}_{\mathrm{int}}=-g\int_{\bf r}\sum_{i;\nu} \sum_{\substack{s\ne s' \\ \xi\ne\xi' }}c_{i;s,\xi}^{\nu,\dagger}({\bf r}) c_{i;s',\xi'}^{\nu,\dagger}({\bf r})  
c_{i;s',\xi'}^{\nu}({\bf r}) c_{i;s,\xi}^{\nu}({\bf r}),~~~~
\end{eqnarray}
where $g>0$ is the coupling strength.
    
\emph{Flat-band stimulated superconductivity}.
Let us first address the Cooper pairing inside the ${\rm AA}$-domains neglecting coupling with their surroundings. Performing the Hubbard-Stratonovich transformation in the Cooper channel for the respective action in the path integral formalism utilizing Hamiltonian $\mathcal{H}$ \cite{supple}, one can introduce the bosonic field $\Delta^{\nu}_{i;s,\xi;s',\xi'}$ for $s \neq s'$ and $\xi \neq \xi'$ inside the $i$-th domain \cite{Gap}. Note that there are several possible combinations of pairing among eight moire-valleys. Although, as long as moire-valley degeneracy is preserved, we shall proceed with a single channel $\Delta_i\equiv \Delta^{\mathrm{A}}_{i;-,-;+,+} $, being insensitive to sublattice index.

\begin{figure}[t!]   
\centering  
\hspace{-.3cm}{ \includegraphics[width=.45\textwidth, height=6cm]{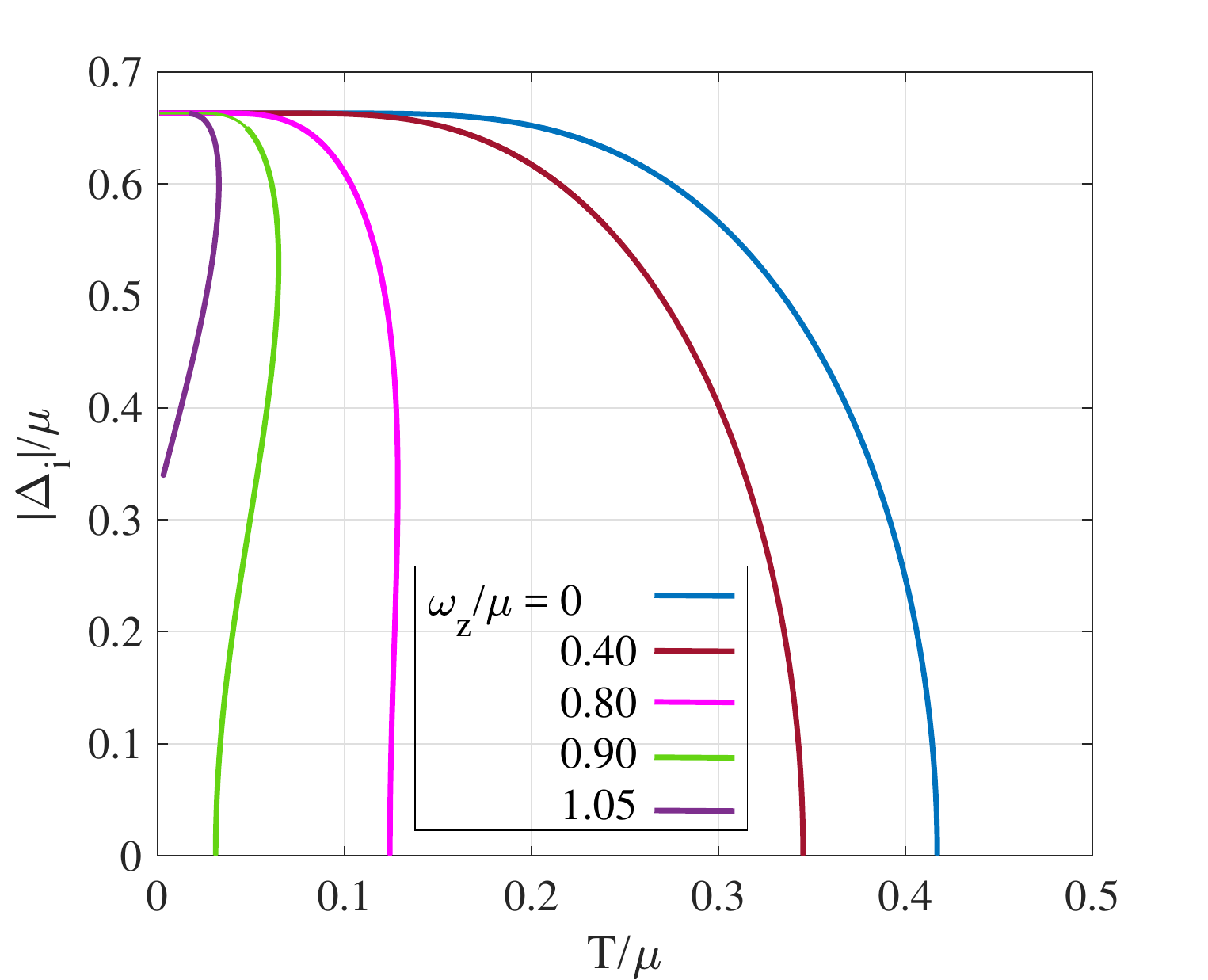}}
\caption{Phase diagram as a function of temperature for different Zeeman energies and fixed interaction strength $g= 1.2g_c$.  Further enhancement of $g$ affects the amplitude of $|\Delta_i|$ without altering the qualitative behaviour.}
\label{direct}
\end{figure}

Hereafter, following the standard approach of integrating out the fermions \cite{supple}, we arrive at the partition function in the static case at temperature $T$ as $Z=\int D(\Delta, \Delta^*)e^{-F[\Delta,\Delta^*]/T}$. The Ginzburg-Landau energy functional is given by $F = \sum_i F_i$ with 
\begin{equation}\label{GL}
F_i= \frac{g_{\mathrm{c}}}{g} \frac{|\Delta_i|^2}{\mu} - 2T\sum_{s=\pm}\ln\left[\cosh\left(\frac{E_{i}-s\omega_z}{2T}\right)\right],
\end{equation}
where $E_{i}=\sqrt{\mu^2+|\Delta_i|^2}$ is the quasiparticle energy and $g_{\mathrm{c}}=2\mu S_{\rm G}$ is the critical interaction strength. The self-consistency equation for $|\Delta_i|$ can be found from (\ref{GL}) as
$E_i/\mu = (g/2g_{\mathrm{c}})\sum_s\tanh\left(E_i-s\omega_z/2T\right)$,
for plot see Fig.~(\ref{direct}). 

For example, 
at $\omega_z=0$ and $T\gg |\Delta_i|$ one finds the transition temperature between metallic and preformed states \cite{Heikkila_meanfield} as
\begin{equation}\label{Tp}
T_{\mathrm{p}}=\frac{\mu}{2 \mathrm{artanh}(g_c/g)}.
\end{equation}
It can be seen that solution exists provided $g>g_c$. It is the strong coupling case.

Note that depending on the parameters ($T, \omega_z, g$), the phase transition between metallic and preformed Cooper pair states can be of the first or second order. 
In particular, at $T=0$ one obtains an equation $E_i/\mu=(g/g_{\mathrm{c}})[1- \theta(|\omega_z|-E_i)]$. For $E_i \geq |\omega_z|$, it has two solutions $|\Delta_i|=\mu\sqrt{(g/g_c)^2-1}$ and $|\Delta_i| = \sqrt{\omega_z^2-\mu^2}$.
The latter exists provided $1 \leq |\omega_z|/\mu \leq g/g_c$.
No solutions exist for $E_i<|\omega_z|$.  
In the case of the first order phase transition, there exists two solutions for $|\Delta_i|$ at a given temperature.
The smaller (larger) value of $|\Delta_i|$ corresponds to the ground (metastable) state.
To illustrate, in the Fig.~(\ref{direct}), the second order phase transition is shown for $\omega_z=(0,0.4)\mu$, whereas the first order phase transition can be seen for $\omega_z=(0.8,0.95,1.05)\mu$. 
   
The saddle point solution $\Delta_i$ is determined up to a random phase on each grain and does not establish global coherence in the system. Although, it might provide a local density of states suppression, which resembles
the pseudogap-like behaviour in the tunneling spectra.
We note that the signatures of the pseudogap in magic-angle TBG above the superconducting transition temperature were seen in the tunneling spectra as reported recently in Ref. \cite{Oh_2021}.

To proceed, in the limit of small gap $\mu \gg |\Delta_i|$, we expand the free energy functional at the vicinity of phase transition in the power series of $|\Delta_i|$ as
\begin{equation}\label{free_energy}
F_i = S_{\rm G}\Big(c_1|\Delta_i|^2+\frac{c_2}{2}|\Delta_i|^4+\frac{c_3}{3}|\Delta_i|^6\Big),
\end{equation} 
where we have dropped $|\Delta_i |$-independent terms which do not play a role here.
The coefficients $c_{1,2,3}$ change sign as a function of ($T, \omega_z, g, g_c$) and determine the stability of preformed Cooper pair state. 
Allowing for arbitrary signs of $c_1$ and $c_2$, we shall restrict $c_3>0$ where expansion Eq. (\ref{free_energy}) is valid. For details see SM \cite{supple}. 
The transition temperature for Cooper pairing instability can be found analyzing the minima of $F_i$ with respect to $\Delta_i$. 
Similar analysis was performed to study the phase transition between homogeneous and LOFF superconducting states \cite{RevModPhys.76.263}. Although, here we neglect spatial variation of $\Delta_i$ within the {\rm AA}-domains.

Let us comment on several limiting cases. At $c_1<0$ and $c_2>0$, there are one maximum at $\Delta_i=0$ and minimum at $|\Delta_i| = \sqrt{|c_1|/c_2}$, which indicates the gap-function in preformed Cooper pair phase. 
Here the line $c_1=0$ defines the second order phase transition. At $c_1>0$ and $c_2<0$, the higher order term $\propto c_3$ is required for the stability of the functional. Here, we have two minima at $\Delta_i = 0$ and $|\Delta_i| = [(|c_2| + \sqrt{c_2^2-4c_1c_3})/(2c_3)]^{1/2}$. In this case there is a first-order phase transition along the line $c_2=-4\sqrt{c_1c_3/3}$ with a discontinuity in the gap $\sqrt{3|c_2|/(4c_3)}$. 
At $c_1<0$ and $c_2<0$, no minima exists and this region is in metallic state. 

We shall now proceed to study the phase coherent superconductivity for which we keep each ${\rm AA}$-domain in preformed state in the regime of $c_1<0$ and $c_2>0$. Note that $c_2<0$ requires Zeeman energy much larger than the one which suppresses the inter-domain coupling, hence we restrict to $c_2>0$ case in what follows.

\begin{figure}[t!]
\begin{tabular}{cc}  
\includegraphics[width=5.0cm, height=1.5cm]{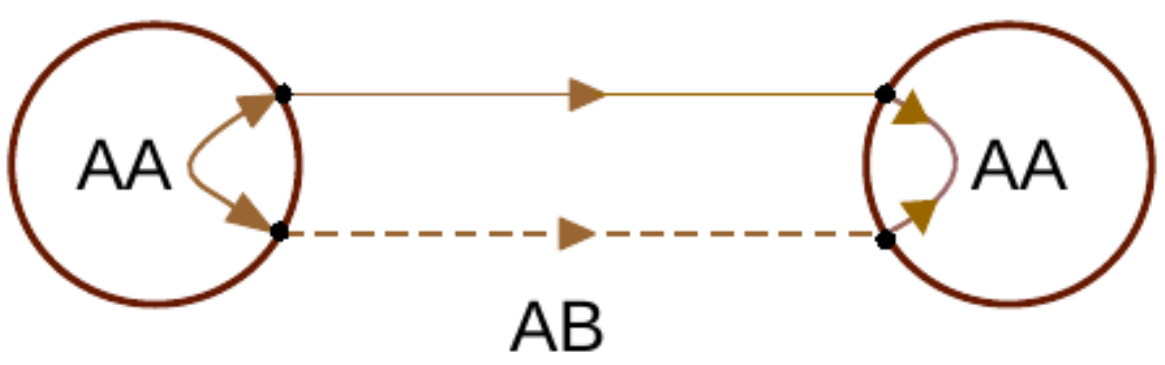}
\end{tabular}
\caption{A schematic sketch of Andreev scattering process between two domains with preformed Cooper pairs.}
\label{bubble}
\end{figure}

\emph{Phase-coherent state: effect of delocalized fermions}.
The delocalized states in the {\rm AB}-regions provide Andreev coupling between different {\rm AA}-domains, Fig. (\ref{bubble}). It might lead the system to reach the global phase-coherence. 
To obtain such coupling, we expand the effective action of the TBG, described by the total Hamiltonian $\mathcal{H}$, up to the fourth power of the hybridization energy between localized and delocalized fermions Eq. \ref{Hybrid}, see SM \cite{supple}. The free energy functional is now given by $F=\sum_i (F_i-\sum_{j\ne i}F_{ij})$  where the coupling between the grains is described by:
\begin{equation}\label{free2}
F_{ij}= 8\pi^2 \tilde{t} \frac{\nu S_{\rm G}^2T}{v_{\mathrm{F}}R_{ij}}
\frac{\cos\left( \frac{2\omega_z}{v_{\mathrm{F}}} R_{ij} \right)}{\sinh\left(\frac{2\pi T }{v_{\mathrm{F}}}R_{ij} \right)}
\Big[\Delta_i^{\ast} \Delta_j
e^{\frac{2ie}{c}\int_{\mathbf{R}_i}^{\mathbf{R}_j}{\bf A} d{\bf R}}+{\rm h.c.} \Big].
\end{equation}   
Here, $v_{\mathrm{F}}=\sqrt{\mu/2m}$ is the Fermi velocity, $\nu=m/2\pi$ is the density of states per spin, valley, and layer of the ${\rm AB}$-region, $R_{ij}=|\mathbf{R}_{i}-\mathbf{R}_j|$ denotes the distance between the grains, and $\tilde{t} = (|t|/\mu)^4/(4\pi)^2\ll 0.1$ is the renormalized tunneling amplitude between two regions. The phase factor attributes to the vector potential of weak magnetic field. The Zeeman field enters through the oscillatory cosine term which drastically affects the coefficients of free energy functional expansion. 

The above coupling term is now treated within mean-field approach assuming weak fluctuations of the order parameter $\Delta_i^\ast\Delta_j\simeq \Delta_i^\ast\la\Delta_j\ra+\la\Delta_i^\ast\ra\Delta_j$. The position dependent mean-field order parameter is defined by
\begin{eqnarray}\label{key}
\la\Delta_i\ra&=& \frac{\int d\Delta d\Delta^{\ast} \Delta_i e^{-F_{\rm MF}/T}}{\int d\Delta d\Delta^{\ast} e^{-F_{\rm MF}/T}}.
\end{eqnarray}
It also determines transition temperature to the phase-coherent superconducting state. Here the mean-field Ginzburg-Landau functional is defined as $F_{\rm MF}=F_i-\sum_{j\ne i} F_{ij} [\Delta_i,\la\Delta_j\ra]$. Expanding the numerator of the above equation and performing functional integration, we arrive at
\begin{eqnarray}\label{SQE_main}
\la\Delta_i\ra &=&
8\pi^2 \tilde{t} \frac{\nu S_{\rm G}^2T}{v_{\mathrm{F}}} \la|\Delta|^2\ra
\\\nonumber
&\times&
\sum_{j\ne i}
\frac{\cos\left(\frac{2\omega_z}{v_{\mathrm{F}}} R_{ij}  \right) 
 }{ \sinh\left(
\frac{2\pi T}{v_{\mathrm{F}}}R_{ij} \right)} \frac{\la\Delta_j\ra}{R_{ij}}
 e^{\frac{2ie}{c}\int_{\mathbf{R}_i}^{\mathbf{R}_j}{\bf A} d{\bf R}},
\end{eqnarray}
where
\begin{equation}\label{Square_MF}
  \la|\Delta|^2\ra=\frac{\int dx x e^{-S_{\rm G}(c_1x+\frac{c_2}{2}x^2)/T}}{\int  dx e^{-S_{\rm G}(c_1x+\frac{c_2}{2}x^2)/T}}.
\end{equation}
Considering the distance between the neighbouring grains to be much smaller than $v_{\mathrm{F}}/\omega_{\mathrm{z}}$ and $v_{\mathrm{F}}/T$, 
we switch to the continuum limit $\sum_{j\ne i}(..)=S_{G}^{-1}\int (..){d \bf R}$, which yields the self consistent equation as
 \begin{align}\label{key4}
 &\left[\beta_0 - \la|\Delta|^2\ra^{-1} +\beta_{1} \mathbf{P}_i^2+\beta_{2} (\mathbf{P}_i^2)^2 \right]\la\Delta_i\ra=0,
\end{align}
where now $\mathbf{P}_i = -i\boldsymbol{\nabla}_{{\bf R}_i}- (2e/c){\bf A}({\bf R}_i)$. The signs of the coefficients $\beta_{0,1,2}$ are sensitive to ($T$, $\omega_z$, $g$, $g_c$), 
and the density of states in the AB-region $\nu$ as shown in SM \cite{supple}. In particular, the coefficient $\beta_1$ changes sign to negative at low temperatures and high magnetic fields, signalling for the inhomogeneous LOFF state ($\beta_1<0$ and $\beta_2>0$). Before proceeding to the LOFF state, let us briefly comment on global-coherence in the case of homogeneous superconductor.
 
\emph{Transition temperature for the homogeneous superconductor}.
Here, we consider the case of homogeneous superconductor in presence of a weak Zeeman field neglecting the out-of-plane magnetic field effect which is strongly suppressed due to small separation between the graphene layers \cite{PhysRevB.101.205116}. Dropping the derivative expansion of the free energy functional in Eq.~(\ref{key4}), we obtain the self consistent equation as $\beta_0\la|\Delta|^2\ra=1$.
Let's analyze this equation in the two limiting cases of weak ($g< g_c$) and strong ($g\gtrsim g_c$) interaction.

At $g< g_c$ and $S_{\rm G}c_1^2/Tc_2\gg1$, we can ignore the $c_2$-term in Eq. \ref{Square_MF} and estimate $ \la|\Delta|^2\ra=T/[2S_{\rm G}(g^{-1}-g_c^{-1})]$. In the absence of Zeeman field, substituting $ \la|\Delta|^2\ra$ into the self-consistency equation (\ref{SQE_main}), we recover the transition temperature between metal and superconductor as
\begin{equation}
 T_c=2\mu\exp\left[-\frac{1}{g\nu\tilde{t}}\left(1-\frac{g}{\g_c}\right)\right].
\end{equation}
This is the BCS-type behaviour except the flat-band induced enhancement of the exponent by a factor $(1-g/g_c)$ \cite{Heikkila_meanfield, PhysRevLett.100.246808}.

It can be shown that the self-consistency equation for the critical Zeeman field ($\omega_{z,c}$) is similar to the BCS one $ \ln |T_c/T|={\rm Re} [\Psi\left(1/2-i\omega_{z,c}/2\pi T\right)]-\Psi\left(1/2\right)$, where $\Psi(z)$ is a psi-function. The $(T,\omega_{z,c})$ phase diagram is shown in the Fig.~(2) of SM \cite{supple}. The above equation can be further simplified at near zero field as $\omega^2_{z,c}=2\epsilon^2 T^2_c(1-T/T_c)$, where $\epsilon=2\pi/\sqrt{14\zeta(3)}\simeq18.4$.

The more interesting situation is realized in the preformed Cooper pair case at $g\gtrsim g_c$. For weak fluctuations $S_{\rm G}c_1^2/c_2T\gg 1$, we have
$ \langle|\Delta|^2\rangle=4\mu^2(1-g_c/g)$. In the absence of the Zeeman field, it yields the transition temperature between the preformed Cooper pair and phase-coherent state
\begin{eqnarray}  
 T_c&=&\pi^2  \mu \left(1-\frac{g_c}{g}\right)\tilde{t} g_c\nu \ln\left|\frac{1}{\tilde{t} g_c \nu}\right|.
\end{eqnarray}
It depends on the density of states of $\mathrm{AB}$-region as $\nu \ln |1/ g_c \nu|$ different to BCS result. We note that the temperature $T_{\mathrm{p}}$ of the Cooper pair formation Eq. (\ref{Tp}) is larger than $T_c$ provided $g_c\nu < 1$ and $\tilde{t}\ll 0.1$. 

Similarly, we obtain the phase transition in presence of the Zeeman field as
$\ln|2\mu/T|-(T/T_c)\ln |2\mu/T_c|={\rm Re}[\Psi\left(1/2-i\omega_{z,c}/2\pi T \right)]-\Psi\left(1/2\right)$, which can be further simplified at the vicinity of zero field as 
$
\omega_{z,c}^2= 2 \epsilon^2 T_c^2(1-T/T_c)\ln\big|2\mu/T_c\big|, 
$
suggesting an enhancement of critical pair-breaking field in the strong coupling case. 
\begin{figure}[t!]
\includegraphics[width=8.0cm]{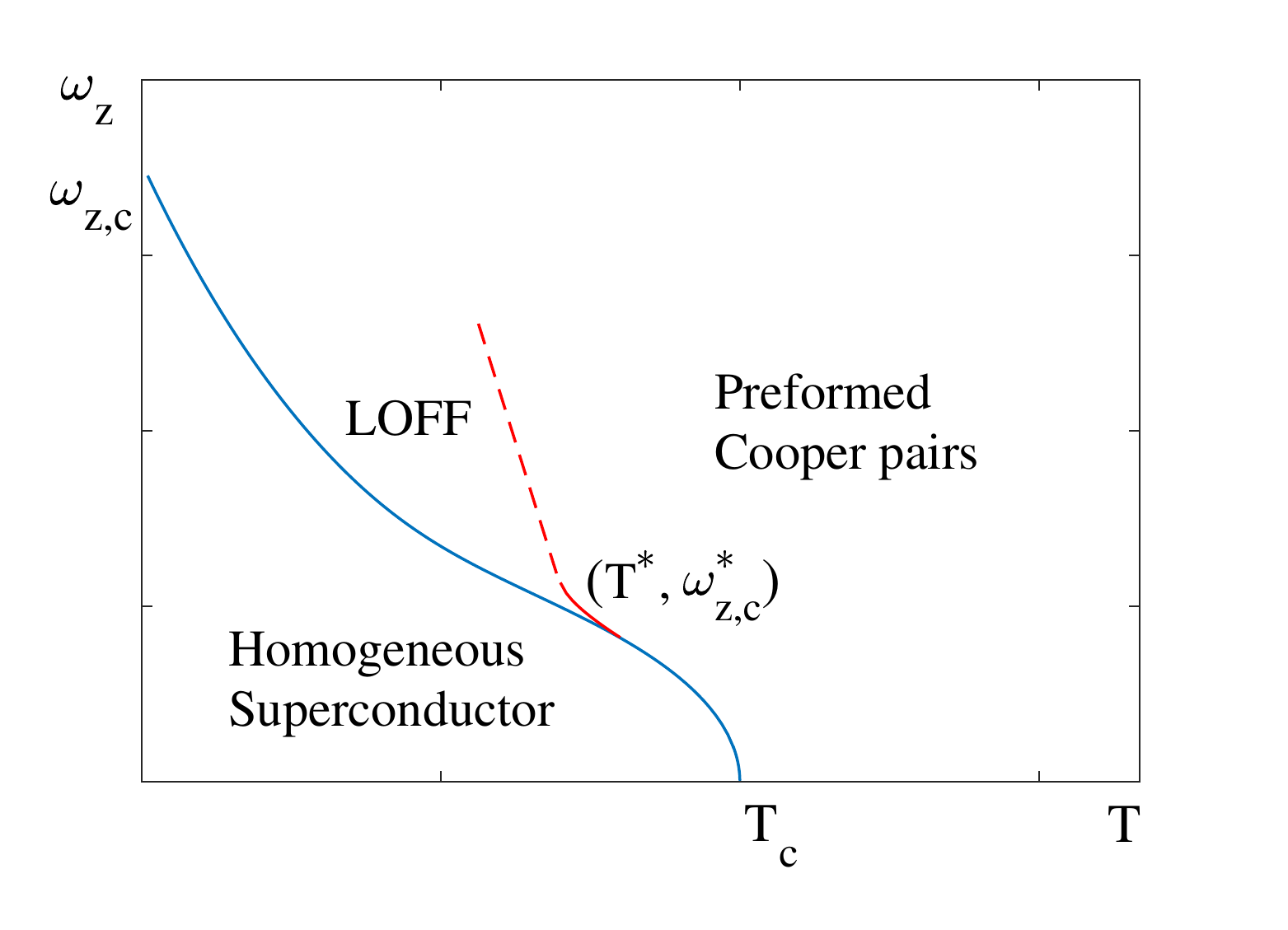}
\caption{A schematic Zeeman energy vs temperature phase diagram in the strong coupling limit. The temperature of the Cooper pair preformation $T_{\rm p}$ is much larger than $T_{c}$ and not shown here.
The dashed line is the extrapolation away from the tricritical point (a meeting point of three phases) around which our self-consistency equation for LOFF state is applicable.}
\label{LOFF_ST}
\end{figure}

\emph{Transition temperature for the LOFF state}.
Let us now discuss the formation of phase-coherent inhomogeneous superconductivity, which originates from the spin-splitting of the conduction electrons by the in-plane magnetic field \cite{Hiroshi_review}.  The Cooper pairing in LOFF state occurs between two spin-split Fermi surfaces with a total  finite momentum ${\bf q}$, which reflects in an oscillatory order parameter as $\Delta_i \propto \Delta\exp(i{\bf q}\cdot {\bf R}_i)$ \cite{Hiroshi_review}. The wavelength of the oscillation is determined by the Zeeman energy.

The self-consistent Eq. (\ref{key4}) transforms to $\beta_0- \la|\Delta|^2\ra^{-1} +\beta_1 q^2+\beta_2 q^4=0$, where $\beta_1<0$ and $\beta_2>0$. It allows to find the critical field for the transition to the LOFF state. However, ${\bf q}$ is still an unknown quantity which can be obtained from the zero supercurrent condition i.e., $\int_{\mathbf{R}}\la {\bf j}\ra = 0$ with
$
 {\bf j}=-4e {\rm Re}\Big\{\Delta^{\ast}_i\left[\beta_1 (-i\boldsymbol{\nabla}_{{\bf R}_i})+2\beta_2(-i \boldsymbol{\nabla}_{{\bf R}_i})^3 \right]\la\Delta_i\ra\Big\}
 $ and $\langle ... \rangle$ standing for the mean-field average as defined in Eq.~(\ref{key}). 
 
Determining $q^2=-\beta_1/2\beta_2$, we find that temperature dependence of the upper critical field for inhomogeneous superconductor can be obtained by solving equation
\begin{eqnarray}\label{LOFF_pd}
I\left(\frac{T}{T_c}\right)&=& {\rm Re}\left[\Psi\left(z\right)\right]-\Psi\left(\frac{1}{2}\right)-\frac{\big\{{\rm Re}\left[\Psi^{(2)}\left(z\right)\right]\big\}^2}{{\rm Re}\left[\Psi^{(4)}\left(z\right)\right]},~~~~~
\end{eqnarray} 
where $z=1/2-i\omega_{z,c}/2\pi T$ introduced for brevity. For weak coupling $g<g_c$ one reproduces the BCS result with $I(x) = \ln|1/x| $, whereas for strong coupling $g \gtrsim g_c$ we obtain
$
I(x) = \ln\big|(2\mu/x T_c)\big|- x \ln\big|(2\mu/T_c)\big|.
$
These two equations describe the ($T,\omega_{z,c}$) phase diagram for an inhomogeneous superconductor, which are bounded by the divergence at $\beta_2=0$ i.e, ${\rm Re}\left[\Psi^{(4)}\left(z\right)\right]=0$. Note that this divergence can be removed only by considering the higher order derivative expansion to the free energy functional.

The LOFF phase takes part in the ($T,\omega_{z,c}$) phase diagram, which is restricted inside the region defined by $\beta_1=0$ and $\beta_1/\beta_2<0$ (see Fig.~2 of Ref.~\cite{supple}). 
All in all, the phase diagram for weak coupling limit resembles the one in usual superconductors \cite{BUZDIN1997341}. 
However, in the strong coupling, the critical Zeeman field is logarithmically enhanced. Contrary to the weak coupling case, we find a phase transition between the preformed and LOFF states as shown on the schematic phase diagram in Fig. (\ref{LOFF_ST}).

\emph{Upper critical magnetic field due to orbital effects}.
Here we briefly comment on the effect of out-of plane critical magnetic field ($H_{c2}$) on the transition to phase-coherent superconducting state ignoring the Zeeman field. Hence, we only consider up to the second order derivative expansion in  Eq. (\ref{key4}), which yields for the weak coupling $H_{c2}= (\Phi_0/2\pi) \left(\epsilon T_{c}/v_F\right)^2\left(1- T/T_{c}\right)$, where $\Phi_0$ is the flux unit $\pi \hbar c/ |e|$. Similarly for the strong coupling
$
H_{c2}= (\Phi_0/2\pi) \left( \epsilon T_{c}/v_F\right)^2 \left(1- T/T_{c}\right)\ln\left|2\mu/T_{c}\right|
$, indicating a logarithmic enhancement of the critical field \cite{Ayu_zyuzin}. We conclude that even though the out-of plane magnetic field does not affect preformed Cooper pair in our model, it enhances the critical field for phase-coherent state in the strong coupling case.
  
 \emph{Berezinskii-Kosterlitz-Thouless (BKT) transition}.
So far we have used the mean-field approach to estimate the transition temperature in TBG. It is noteworthy to mention that in 2D systems such an approach lacks accuracy because of the absence of usual long-range correlations. The superconducting phase transition in 2D system might be described by the formation of quasi-long-range bound pairs of vortices with opposite vorticity. 
The phase transition occurs when such bound vortex pairs are broken and become free. 
Thus our mean field calculation gives the upper bound for the phase transition as in the case of monolayer graphene \cite{PhysRevLett.100.246808}. The BKT transition for the LOFF state might be an interesting problem and will be considered separetely.
  
\emph{Summary}.
We have presented a theoretical investigation of the flat-band induced preformed Cooper pairing in magic angle TBG. We have developed an approach within the mean-field approximation to investigate the superconducting transition temperature. 
The TBG might be a playground for an unusual phase transition between preformed Cooper pair state and inhomogeneous LOFF state.

\emph{Acknowledgements}.
This work is supported by the Academy of Finland Grant No. 308339.  Authors acknowledge Jose Lado for useful comments. 

\bibliography{Bib_twisted_arxiv}

\begin{thebibliography}{35}
\bibitem [{Gap()}]{Gap}%
  \BibitemOpen
  \href@noop {} {}\bibinfo {note} {{Within the BCS formalism the gap function
  is defined as $\Delta^{\nu}_{i;s,\xi;s',\xi'}(\mathbf{r}) = g\la
  c_{i;s,\xi}^\nu(\mathbf{r}) c_{i;s',\xi'}^\nu(\mathbf{r})\ra$}}\BibitemShut
  {NoStop}%
\end{thebibliography}

 \end{document}